\documentclass[aps,nofootinbib,twocolumn,floatfix,superscriptaddress,secnumarabic]{revtex4-1}
\usepackage{amsfonts,amssymb,epsfig,verbatim,mathbbol,amstext,amsmath,psfrag}
\usepackage{graphicx}
\usepackage{t1enc}
\usepackage{amsmath}
\usepackage{subfigure}
\usepackage{dsfont}
\usepackage{slashed}

\usepackage{colordvi}
\usepackage{xcolor}
\usepackage{color}

\newif\ifSMver
\SMverfalse

\ifSMver
\newcommand{\mysec}[1]{{\em #1} ---}

\newcommand{\appendixsection}{%
\setcounter{equation}{0}
\setcounter{section}{0}
\renewcommand{\theequation}{S\arabic{equation}}
\onecolumngrid
\vspace*{.7cm}
\hrule
\vspace*{.04cm}
\hrule
\begin{center}
\vspace*{.4cm}
{\bf \large Supplemental Material}
\vspace*{.5cm}
\end{center}
\twocolumngrid
}
\else
\newcommand{\mysec}[1]{\section{#1}}

\newcommand{\appendixsection}{\appendix}
\fi

\newcommand{\be}{\begin{equation}}
\newcommand{\ee}{\end{equation}}

\newcommand{\dd}{\textmd{d}}
\newcommand{\Tr}{\textmd{Tr}}
\newcommand{\Z}{\mathcal{Z}}
\newcommand{\D}{\mathcal{D}}
\newcommand{\expv}[1]{\left \langle #1 \right \rangle}

\newcommand{\Bielefeld}{Fakult\"at f\"ur Physik, Universit\"at Bielefeld, D-33615 Bielefeld, Germany.}

\usepackage[    bookmarks,
                 bookmarksopen = true,
                 bookmarksnumbered = true,
                 breaklinks = true,
                 linktocpage,
                 colorlinks = true,
                 linkcolor = blue,
                 urlcolor  = blue,
                 citecolor = blue,
                 anchorcolor = green,
                 hyperindex = true,
                 hyperfigures]
		{hyperref}        
		
\begin{document}

\title{
QCD phase diagram and equation of state in background electric fields
}

\author{G.~Endr\H{o}di,}
\author{G.~Mark\'o}
\affiliation{\Bielefeld}

\begin{abstract}
The phase diagram and the equation of state of QCD is investigated in the presence of 
weak background electric fields
by means of continuum extrapolated lattice simulations. 
The complex action problem at nonzero electric field 
is circumvented by a novel Taylor expansion, enabling the determination of the linear response of the thermal QCD medium to constant electric fields -- in contrast to simulations at imaginary electric fields, which, as we demonstrate, involve an infrared singularity.
Besides the electric susceptibility of QCD matter, we determine the dependence of the Polyakov loop on the 
field strength to leading order. 
Our results indicate a plasma-type behavior with a negative susceptibility at all temperatures, as well as an increase in the 
transition temperature
as the electric field grows.
\end{abstract}


\maketitle

\mysec{Introduction} 
The phase structure of Quantum Chromodynamics (QCD) in the presence
of background electromagnetic fields is an essential attribute of the fundamental theory 
of quarks and gluons and, accordingly, a subject of active
theoretical research. 
The electromagnetic response of the QCD medium is relevant for 
a range of physical situations, e.g.\ the phenomenology of 
heavy-ion collisions, the description of neutron star interiors
or the evolution of our universe in its early stages, see the reviews~\cite{Kharzeev:2013jha,Miransky:2015ava}.
If in these settings the electromagnetic fields are sufficiently long-lived compared to the strong scale, it is appropriate to consider QCD matter in a background magnetic or electric field in equilibrium. 

Before equilibration, electric fields $E$ induce a dynamical response via the electrical conductivity of the medium~\cite{Meyer:2011gj}. The subsequently emerging equilibrium necessarily involves -- in contrast to the case of magnetic fields $B$ --
an inhomogeneous charge distribution $n(x)$ in the thermal medium while having constant temperature $T$ everywhere~\cite{landau2013statistical}. The distribution is uniquely fixed by the requirement that pressure gradients and electric forces cancel each other and thus no currents flow~\cite{Luttinger:1964zz}.
The equilibrium system is therefore described by a {\it local canonical} statistical ensemble, where $n(x)$ is held fixed. It differs from the grand canonical ensemble parameterized by chemical potentials, employed usually at $E=0$.
This aspect renders comparisons between equilibrium systems at $E>0$ and $E=0$, e.g.\ by means of lattice simulations, 
problematic.  

Moreover, the proper definition of the equilibrium state at $E>0$ requires infrared regularization (e.g.\ a finite spatial volume $V$) that prevents charges to be accelerated to infinity. As we have demonstrated recently within perturbative QED~\cite{Endrodi:2022wym}, the $E\to0$ and $V\to\infty$ limits of this setup do not commute at nonzero temperature. This renders approaches based on Schwinger's exact $E>0$ infinite-volume propagator~\cite{Schwinger:1951nm} and 
infrared-regularized weak-field expansions in the manner of Weldon~\cite{Weldon:1982aq} inherently different. 
For a certain physical setting, the boundary conditions determine which is the appropriate limit to consider.
The generalization of these ideas to the case of QCD enables one to explore the impact of background electric fields on strongly interacting matter as well as the associated phase diagram: our objectives in the present letter.

The impact of magnetic fields on the QCD crossover~\cite{Aoki:2006we,Bhattacharya:2014ara} and the corresponding phase diagram is well understood and 
has been studied extensively on the lattice~\cite{DElia:2010abb,Bali:2011qj,Bali:2012zg,Endrodi:2015oba,DElia:2021yvk}, as well as within models and effective theory approaches (for a recent review, see Ref.~\cite{Andersen:2014xxa}). In contrast, electric fields render the QCD action complex, hindering standard lattice simulations.
Alternative approaches include Taylor-expansions~\cite{Fiebig:1988en,Christensen:2004ca,Engelhardt:2007ub,Lee:2023rmz}, calculations at imaginary electric fields~\cite{Detmold:2009dx,Detmold:2010ts,Lujan:2014kia,Freeman:2014kka,Lujan:2016ffj,Yang:2022zob} and simulations with electric fields that couple to the isospin charge of quarks~\cite{Yamamoto:2012bd}. Still, there are no existing results for the QCD equation of state nor the phase diagram. The latter has only been studied within effective theories like the linear $\sigma$ model~\cite{Suganuma:1990nn}, variants of the Nambu-Jona-Lasinio (NJL) model~\cite{Klevansky:1989vi,Babansky:1997zh,Tavares:2019mvq,Tavares:2023ybt} and the Euler-Heisenberg effective action~\cite{Ozaki:2015yja}. These calculations are all based on the Schwinger propagator.

In this letter, we determine the QCD equation of state  and the phase diagram on the lattice for the first time for weak background electric fields.
The complex action problem is circumvented via a Taylor-expansion: this corresponds to the Weldon-type regularization of the electrically polarized thermal medium and is the proper description of a finite system, where equilibration takes place in the presence of a weak electric field.
The expansion is based on the method we developed in Refs.~\cite{Endrodi:2021qxz,Endrodi:2022wym}, and resembles the analogous approach for background magnetic fields~\cite{Bali:2015msa,Bali:2020bcn,Buividovich:2021fsa}.
Besides the leading coefficient -- the electric susceptibility 
of QCD matter -- we also determine the leading series of the Polyakov loop. 
Using this observable, we construct the phase diagram and demonstrate that 
the transition temperature increases as $E$ grows -- contrary to existing model predictions, e.g.~\cite{Tavares:2019mvq}. Finally, we demonstrate that lattice simulations at nonzero imaginary electric fields cannot be used to directly calculate the electric susceptibility due to the singular change of ensembles between $E=0$ and $iE\neq0$.
Some of our preliminary results have already been presented in Ref.~\cite{Endrodi:2021qxz}.

\mysec{Lattice setup}
QCD matter in thermal equilibrium is a medium that can be polarized by weak background electromagnetic fields. The associated static linear response is characterized by the electric and magnetic susceptibilities (we employ the same notation as in Ref.~\cite{Endrodi:2022wym}). 
These are defined via the matter free energy density $f$,
\be
\xi_b=-\left.\frac{\dd^2 f}{\dd (eE)^2}\right|_{E=0}\,, \quad\quad
\chi_b=-\left.\frac{\dd^2 f}{\dd (eB)^2}\right|_{B=0}\,.
\label{eq:chidef}
\ee
Here, the subscript $b$ indicates that both susceptibilities contain ultraviolet divergent terms that must be 
subtracted via additive renormalization, see below.
The elementary charge $e$ is included so that we can work with the renormalization group invariants $eE$ and $eB$.

The matter free energy density can be rewritten using the partition function $\Z$ of the system. 
Using the rooted staggered formalism of lattice QCD, it is given by the Euclidean path integral over the gluon links $U$,
\be
\Z= \int \D U\, e^{-\beta S_g} \prod_f \det [\slashed{D}(q_f)+m_f]^{1/4}\,,
\label{eq:pathint}
\ee
where $\beta=6/g^2$ is the inverse gauge coupling and $m_f$ denotes the quark 
masses with $f=u,d,s$ running over the quark flavors.
The simulations are done in a periodic spatial volume $V=L^3$ with linear size $L$. Note that $\Z$ corresponds to the grand canonical ensemble; its relation to the canonical one at $E>0$ is discussed in App.~\ref{app:ploop}.
In Eq.~\eqref{eq:pathint}, $S_g$ is the gluon action (in our discretization, the tree-level improved Symanzik 
action) and $\slashed{D}_f$ is the staggered Dirac operator (including a twofold stout smearing of the links) that contains the quark charges $q_u/2=-q_d=-q_s=e/3$. The quark masses are set to their physical values as a function of the 
lattice spacing $a$~\cite{Borsanyi:2010cj}. Further details of the action and of our simulation 
algorithm are given in Refs.~\cite{Aoki:2005vt,Bali:2011qj}.

The electromagnetic vector potential $A_\nu$
enters the Dirac operator in the form of temporal parallel transporters
$u_{\nu,f}=\exp(iaq_f A_\nu)$ multiplying the gluon links $U_\nu$.
We choose a gauge where $A_0(x_1)$ represents the electric field and $A_2(x_1)$ the magnetic field (both pointing in the $x_3$ direction).
While magnetic fields are identical in Minkowski and Euclidean space-times, the vector potential relevant for the electric field undergoes a Wick rotation so that $A_4=iA_0$, similarly to the case of a chemical potential $\mu$.
Finally we mention that in our setup, quarks do not couple to dynamical photons but only to the 
external gauge field. 
The independent thermodynamic 
variable is the field $E$ that enters the Dirac operator, analogously to the situation for magnetic fields~\cite{Bonati:2013vba}.

\mysec{Observables}
As we demonstrated in Ref.~\cite{Endrodi:2022wym}, the susceptibilities of Eq.~\eqref{eq:chidef} are related to derivatives of the electromagnetic vacuum polarization tensor with respect to spatial momenta. For our gauge choice, these relations read in terms the Euclidean polarization tensor $\Pi_{\mu\nu}$,
\be
\xi_b = -\frac{1}{2}\left.\frac{\partial^2 \Pi_{44}(k)}{\partial k_1^2}\right|_{k=0}, \quad
\chi_b = \frac{1}{2}\left.\frac{\partial^2 \Pi_{22}(k)}{\partial k_1^2}\right|_{k=0},
\label{eq:chimeas}
\ee
with a spatial momentum $k=(k_1,0,0,0)$. In other words, the zero momentum limit is considered at vanishing time-like frequency, reflecting the static nature of the susceptibilities. The negative sign for $\xi_b$ in~\eqref{eq:chimeas} appears due to the Wick rotation of the electric field. 
We highlight that the equilibrium systems at different values of $E$ exhibit different charge profiles $n(x_1)$, and this implicit $E$-dependence is taken into account properly in Eq.~\eqref{eq:chimeas} for the calculation of $\xi_b$~\cite{Endrodi:2022wym}. In fact, without this contribution, $\xi_b$ would diverge in the $k_1\to0$ limit.

The vacuum polarization tensor is defined as the correlator
\be
\Pi_{\mu\nu}(k)=\int \dd^4 x\, e^{ikx} \expv{j_\mu(x) j_\nu(0)}\,,
\ee
of the electromagnetic current
$j_\mu = \sum_f \frac{q_f}{e} \bar\psi_f \gamma_\mu \psi_f$,
for which we use the conserved (one-link) staggered vector current.
It is convenient to evaluate~\eqref{eq:chimeas} in coordinate space, where the bare susceptibilities become~\cite{Endrodi:2021qxz,Endrodi:2022wym}
\be
\xi_b = -\big\langle G_{44}^{(2)}\big\rangle, \qquad
\chi_b = \big\langle G_{22}^{(2)}\big\rangle, 
\label{eq:xibchib}
\ee
containing the second moment of a partially zero-momentum projected two-point function
\begin{align}
G_{\mu\nu}^{(2)} &= \int_0^{L/2} \!\!\dd x_1 \,x_1^2 \,G_{\mu\nu}(x_1), \label{eq:chicalc}\\
G_{\mu\nu}(x_1)&=\int \dd x_2\, \dd x_3\, \dd x_4 \,\; j_\mu(x) j_\nu(0)\,.
\label{eq:Gdef}
\end{align}
The Grassmann integral over quark fields is understood to be implicitly carried out on the right hand side of the last equation. 

Both susceptibilities undergo additive renormalization. This originates 
from the multiplicative divergence in the electric charge $e$~\cite{Dunne:2004nc,Bali:2014kia,Bali:2020bcn}. Being 
temperature-independent, the divergence cancels in
\be
\xi = \xi_b(T)-\xi_b(T=0)\,,\quad
\chi = \chi_b(T)-\chi_b(T=0)\,,
\label{eq:xirdef}
\ee
which sets $\xi=\chi=0$ at zero temperature.
In fact, at $T=0$ Lorentz invariance ensures that $\big\langle G_{22}^{(2)}\big\rangle =
\big\langle G_{44}^{(2)}\big\rangle$, implying that the 
bare magnetic and electric susceptibilities coincide up to a minus sign. 
To renormalize the electric susceptibility, we can therefore employ the 
existing results for $\chi_b(T=0)$ from Ref.~\cite{Bali:2020bcn}.

Next we consider the bare Polyakov loop operator,
\be
P_b =
\frac{1}{V}\int\dd^3\mathbf{x}\;\textmd{Re}\, \Tr\, \prod_{x_4} U_4(x)\,.
\ee
Its expectation value is related to the 
free energy of a static, electrically neutral color charge and is often taken 
as a measure of deconfinement. Just as for $\xi_b$, the contribution of the equilibrium charge profile needs to be taken into account for the $E$-dependence of the Polyakov loop as well.
As we show in App.~\ref{app:ploop},
the proper second-order expansion of $\expv{P_b}$ is given by the correlator
\be
\varphi_E^n \equiv \left.\frac{\dd^2 \!\expv{P_b}^n}{\dd (eE)^2}\right|_{E=0} = \frac{V}{T}\left[
-\big\langle P_b \,G_{44}^{(2)}\big\rangle+\big\langle P_b\big\rangle\big\langle G_{44}^{(2)}\big\rangle \right]\,,
\label{eq:Pcalc}
\ee
where the superscript $n$ on the left denotes that the derivative is evaluated along the equilibrium condition specified by the local charge profiles.
Analogously, the magnetic derivative of $\expv{P_b}$ can be obtained by replacing $-G_{44}^{(2)}$ 
by $G_{22}^{(2)}$ in Eq.~\eqref{eq:Pcalc}, although in that case nontrivial charge distributions do not appear.

The bare Polyakov loop is subject to multiplicative renormalization~\cite{Borsanyi:2012uq},
\be
P(a,T,E)= P_b(a,T,E)\cdot
\left(\frac{P_\star}{P_b(a,T_\star,E=0)}\right)^{T_\star/T}\,,
\label{eq:Prdef}
\ee
where the renormalization factor is independent of the background field
and has been determined for our lattice spacings in Ref.~\cite{Bruckmann:2013oba}. This renormalization 
fixes $P=P_\star$ at $T=T_\star$ and $E=B=0$.
In our renormalization 
scheme we choose $T_\star=162\textmd{ MeV}$ and $P_\star=1$.

\begin{figure}[t]
 \centering
 \includegraphics[width=8cm]{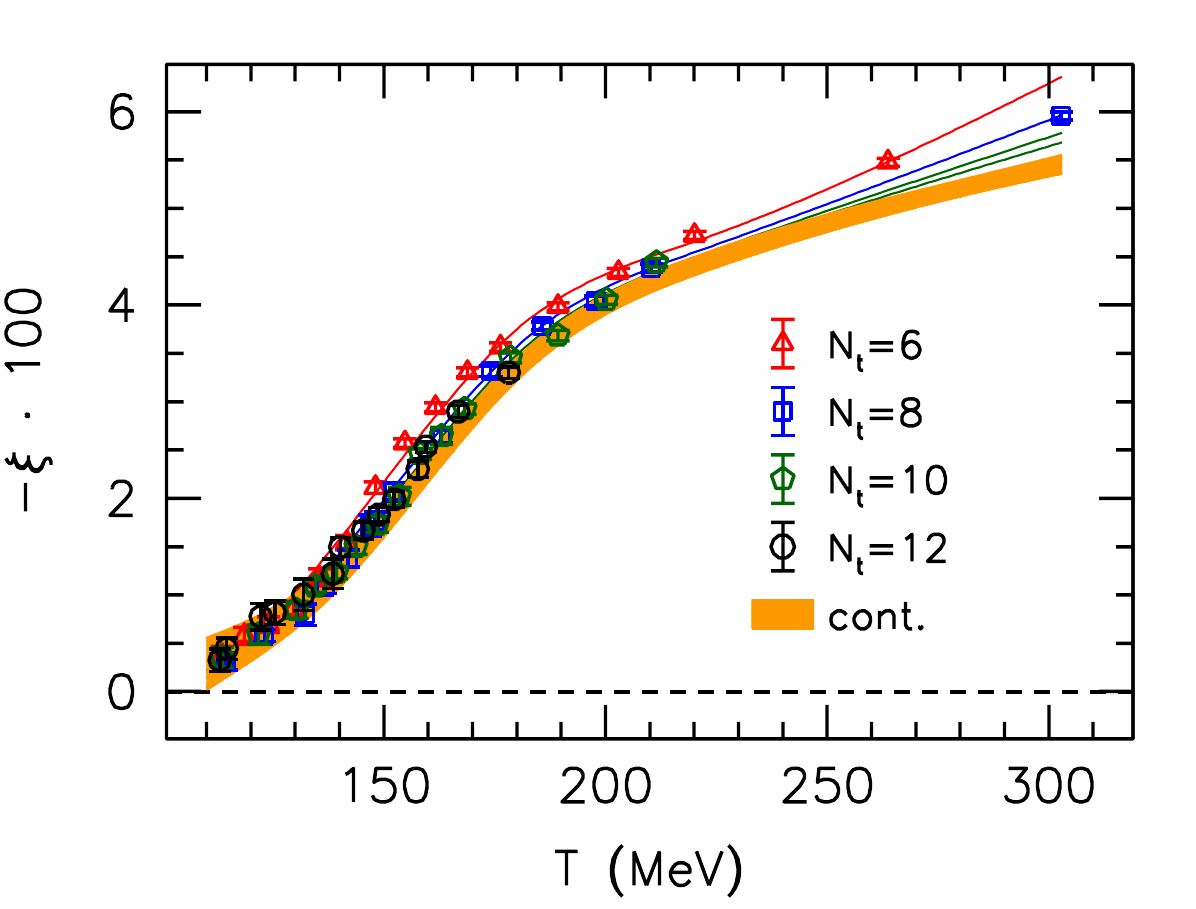}
 \includegraphics[width=8cm]{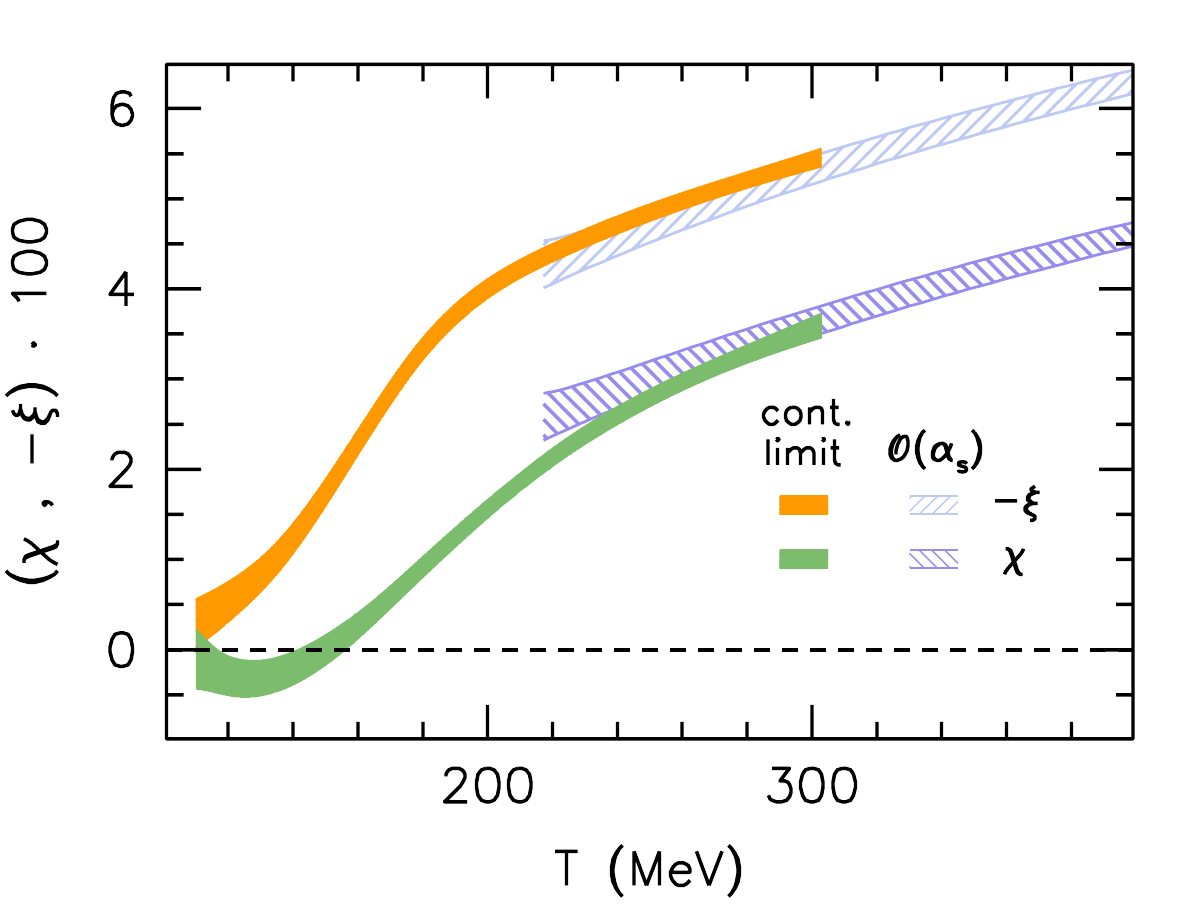}
 \caption{\label{fig:susc_comp}Upper panel: the negative of the renormalized electric susceptibility as a function 
 of the temperature for four lattice spacings (colored symbols) and a continuum 
 extrapolation (orange band). Lower panel: continuum extrapolated magnetic (green) and electric (orange) susceptibilities (solid) compared to leading-order perturbation theory (dashed).}
\end{figure}

\mysec{Results: susceptibility}
We have measured the zero-momentum projected correlator~\eqref{eq:Gdef} for a broad range of temperatures on $N_t=6,8,10$ and $12$ lattice ensembles. Finite volume effects were checked using $16^3\times6$ and $24^3\times6$ lattices. We report on the details of the measurements and the analysis in App.~\ref{app:corr}.
The correlator is convolved with the quadratic kernel according to Eq.~\eqref{eq:chicalc} to find the
bare electric susceptibility $\xi_b$, and its renormalization~(\ref{eq:xirdef}) is carried out by subtracting the zero-temperature contribution. 

The negative of the so obtained $\xi$ is plotted in the upper panel of Fig.~\ref{fig:susc_comp}. 
A continuum extrapolation is performed via a multi-spline fit of all data points,
taking into account $\mathcal{O}(a^2)$ lattice artefacts. The systematic error 
of the fit is estimated by varying the spline node points and including $\mathcal{O}(a^4)$ discretization errors in the fit at low temperatures. For all temperatures 
we observe $\xi<0$, translating to an electric permittivity below unity -- a characteristic feature of plasmas~\cite{jackson_classical_1999}.
At high $T$, our results may be compared to the high-temperature limit calculated for
non-interacting
quarks of mass $m$~\cite{Endrodi:2022wym},
\be
\xi^{\rm free}\xrightarrow{T\to\infty} -\sum_f (q_f/e)^2 \frac{N_c}{12\pi^2} \cdot \left[ \log \frac{T^2\pi^2}{m^2}-2\gamma_E - 1 \right]\,,
\label{eq:pertform}
\ee
where $N_c=3$ is the number of colors~\cite{footnote1}. 
In full QCD,
the quark mass is replaced by a QED renormalization scale $\mu_{\rm QED}$ that can be determined at $T=0$ and is found to be $\mu_{\rm QED}=115(6) \textmd{ MeV}$~\cite{Bali:2020bcn}, close to the mass of the lightest charged hadron i.e.\ the pion. 
Moreover, QCD corrections are included by taking into account $\mathcal{O}(\alpha_s)$ effects in the prefactor, the QED $\beta$-function~\cite{Baikov:2016tgj,Bali:2020bcn}. The associated thermal scale is varied between $\pi T$ and $4\pi T$ for error estimation.
The so obtained curve lies very close to our results at high temperature, as visible in the lower panel of Fig.~\ref{fig:susc_comp}, where we also show the corresponding results for $\chi$ from Ref.~\cite{Bali:2020bcn}.

\begin{figure}[t]
 \centering
 \includegraphics[width=8cm]{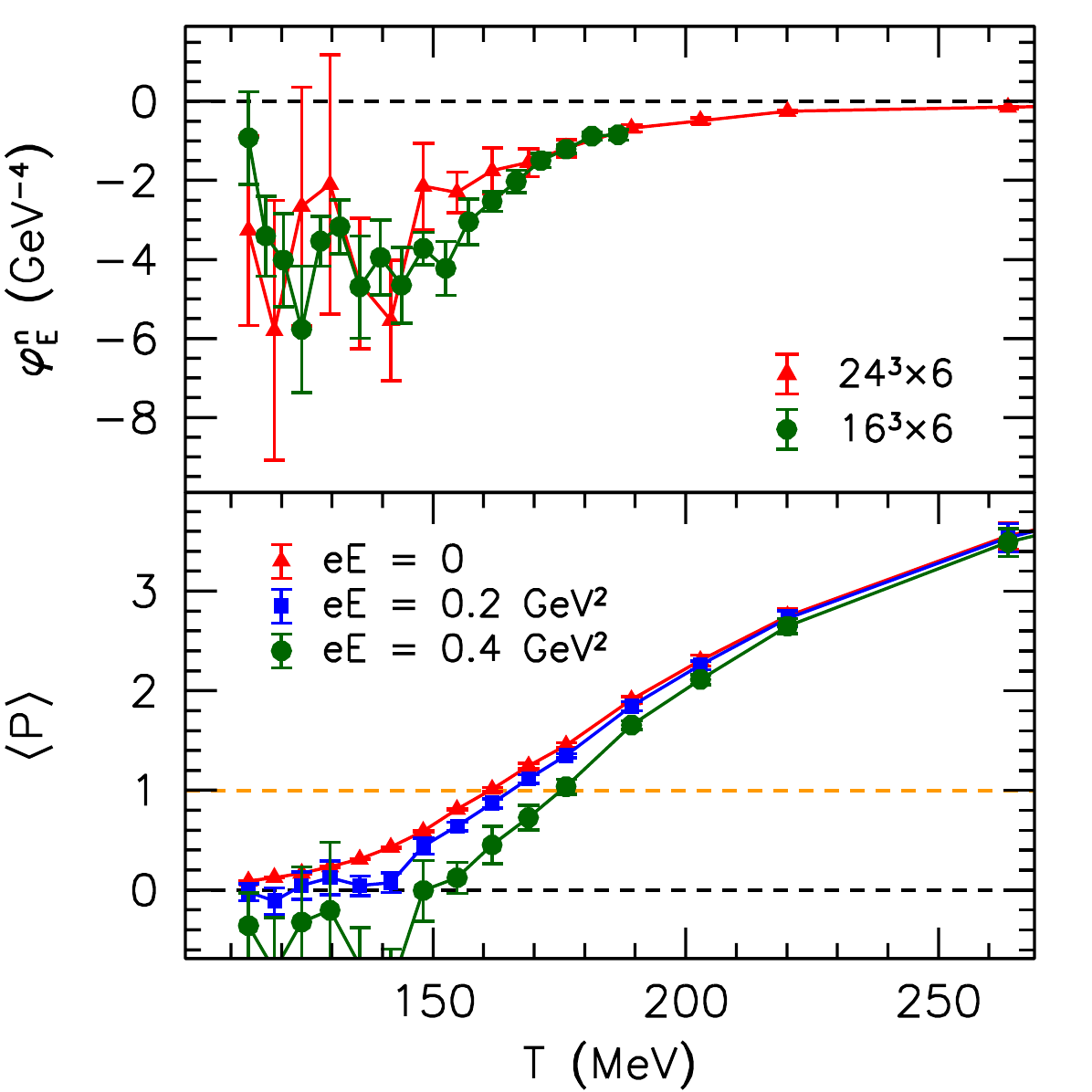} \caption{\label{fig:ploopE2}Upper panel: the leading expansion coefficient of the bare Polyakov 
 loop as a function of the temperature as obtained on our $24^3\times6$ (red) and $16^3\times6$ (green) ensembles. Lower panel: renormalized Polyakov loop at nonzero electric 
 fields, constructed from the leading Taylor series. 
 The crossing point with the dashed yellow line is identified 
 with $T_c$. }
\end{figure}

\mysec{Results: phase diagram}
Next we turn to the Polyakov loop. 
Its leading expansion is given by Eq.~\eqref{eq:Pcalc}, containing the correlator of the bare observable with $-G_{44}^{(2)}$. This quantity is plotted in the upper
panel of 
Fig.~\ref{fig:ploopE2} for our $N_t=6$ lattices, revealing negative values for the 
complete range of temperatures, i.e.\ a reduction of the Polyakov loop by the electric 
field. Finite volume effects are found to be small, although the results at low temperature have large statistical uncertainties. 
Using the results for the Polyakov loop at $E=0$~\cite{Bruckmann:2013oba} and the multiplicative renormalization factor from Eq.~\eqref{eq:Prdef}, we construct the $E$-dependence of 
$\expv{P}$, see the lower panel of Fig.~\ref{fig:ploopE2}. 
The Polyakov loop is known to exhibit a smooth temperature-dependence, so 
that a precise determination of its inflection point is cumbersome already at $E=0$.
As an alternative, we associate the 
transition temperature $T_c$ with the point where $\expv{P}=1$ holds. Defined 
in this manner, the lower panel of Fig.~\ref{fig:ploopE2} clearly shows that $T_c$ is increased by $E$.

\begin{figure}[t]
 \centering
 \includegraphics[width=8cm]{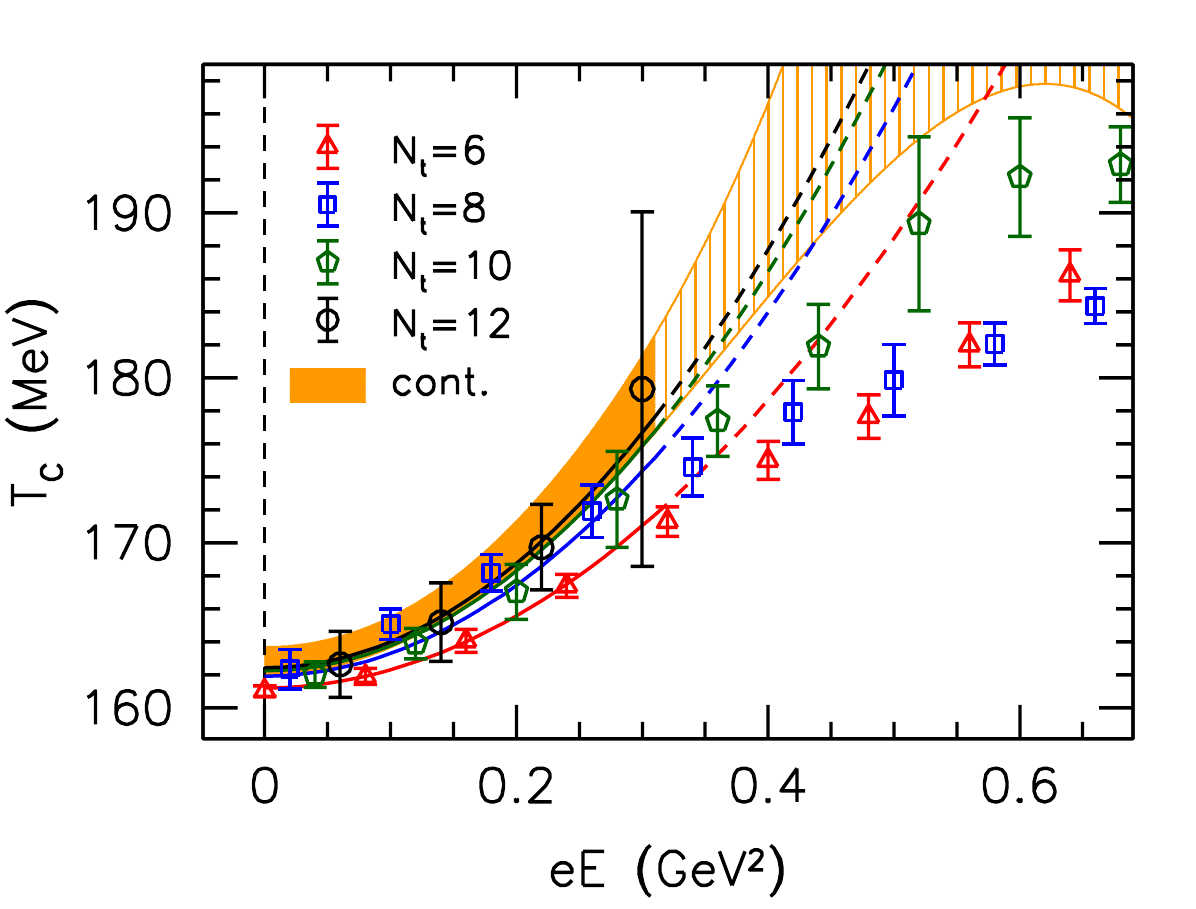}
 \caption{\label{fig:phasediag}Transition temperature as a function of 
 the electric field for different lattice spacings (colored symbols)
 and a continuum extrapolation (yellow band). Higher-order effects in $eE$ become non-negligible for $eE\gtrsim 0.3\textmd{ GeV}^2$, indicated by the dashed section of the fits.}
\end{figure}

We repeat this analysis for all four lattice spacings. Our results for the transition 
temperature are shown in Fig.~\ref{fig:phasediag}, confirming the significant 
enhancement of $T_c$ as the electric field grows. 
We perform the continuum extrapolation by a quadratic fit of $T_c(E)$ 
taking into account $\mathcal{O}(a^2)$ 
lattice artefacts. To estimate the systematic error, we vary the fit range and also allow a quartic term in the fit. The fits are found to be stable for the region $E\lesssim 0.3 \textmd{ GeV}^2$~\cite{footnote2}.
The curvature of the transition line is found to be
\be
\kappa_E \equiv \left.\frac{\partial^2 T_c(E)}{\partial (eE)^2}\right|_{E=0} = 
0.37(9)\textmd{ GeV}^{-3}\,.
\ee
Furthermore, we find the transition to get stronger as $E$ grows, 
revealed by an enhancement of the slope of the Polyakov loop as a function 
of $T$, see Fig.~\ref{fig:ploopE2}. However, due to the large uncertainties 
at low temperatures, we cannot make a quantitative 
statement about this aspect.

\mysec{Results: imaginary electric fields}
Finally, we consider lattice simulations at constant imaginary electric fields $iE$.
In a finite periodic volume at nonzero temperature, the allowed field values are quantized as $ieE = 6\pi T/L \cdot N_e$ with the `flux' quantum $N_e\in\mathds{Z}$~\cite{tHooft:1979rtg}. 
This setup does not correspond to the analytic continuation of the local canonical ensemble 
as described in the introduction. Nevertheless, it involves a global constraint: the total electric charge in the periodic volume vanishes. As a consequence, this setup is independent of the global imaginary chemical potential $i\mu$. Indeed, including any $i\mu\neq0$ can be canceled in the gauge field by a mere coordinate translation $x_1\to x_1-\mu/E$. 
This is in stark contrast to the situation at $E=0$, where a dependence on $i\mu$ is naturally present.

\begin{figure}[t]
 \centering
 \hspace*{-.1cm}
 \includegraphics[width=9cm]{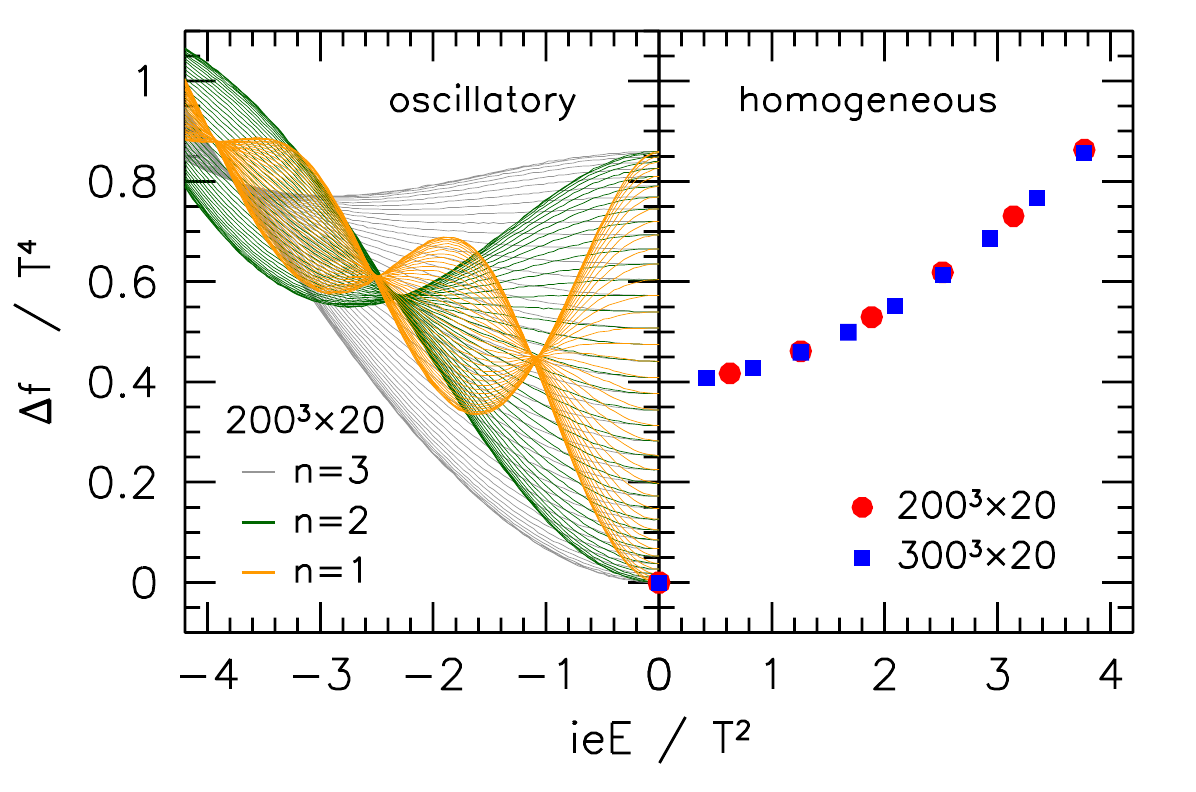}
 \caption{\label{fig:logdet}Free energy density as a function of homogeneous (right side) and oscillatory (left side) imaginary electric fields. The results for different imaginary chemical potentials correspond to the set of curves in the left ($i\mu$ grows from $0$ to $\pi T$ from the bottom to the top), while they lie on top of each other on the right.}
\end{figure}

To discuss this issue, we neglect gluonic interactions in the following. In this simplified setting, we can calculate the free energy density directly via exact diagonalization of the Dirac operator~\cite{footnote3}. In the right side of Fig.~\ref{fig:logdet} we show the results for $\Delta f=f-f(E=\mu=0)$ obtained on a $200^3\times20$ lattice with quark mass $m/T=0.08$. 
As expected, $f$ is found to be independent of the imaginary chemical potential in the whole range $0\le i\mu \le \pi T$ at any $iE\neq0$. 
The comparison to a larger volume $300^3\times20$ shows that the smallest allowed electric field value approaches zero in the thermodynamic limit, but $\lim_{iE\to0} f(iE)\neq f(iE=0)$. Instead, the data points rather accumulate towards the average of $f$ over all possible imaginary chemical potential values -- i.e.\ a canonical setup where the total charge is constrained to zero.
Altogether, we conclude that the dependence of $f$ on $iE$ is singular at $E=0$ in the thermodynamic limit, rendering simulations with homogeneous imaginary electric fields unsuited for the evaluation of $\xi$.

In addition, the left side of Fig.~\ref{fig:logdet} shows $\Delta f$ for oscillatory imaginary electric fields with the profile $E(x_1)= E\sqrt{2} \cos (2\pi n x/L)$. In this case, the role of the infrared regulator is played by the wave number $n$ and not by the volume. Moreover, here $E$ is a continuous variable but $n\in\mathds{Z}$ is discrete. This setup does not fix the overall charge and, therefore, maintains the dependence of $f$ on $i\mu$. Indeed, the results reveal a continuous behavior as a function of $iE$ and $i\mu$. 
However, as visible in the plot, the results again approach a singular behavior as the infrared regulator is removed: the curves collapse to a set of $i\mu$-independent nodepoints approaching the $iE=0$ axis. In particular, the curvature of $f$ with respect to $iE$ diverges for $n\to0$. 
Thus, the homogeneous limit of the setup with oscillatory imaginary fields reproduces what we have already seen for the homogeneous case.

\mysec{Discussion}
In this letter we studied the thermodynamics of QCD at nonzero background electric 
fields $E$ via lattice simulations with physical quark masses.
To avoid the complex action problem at $E>0$, we employed a 
leading-order Taylor-expansion.
This approach is more complicated than the analogous expansion in a chemical potential, because the impact of $E$ on the equilibrium charge distribution needs to be taken into account~\cite{Endrodi:2022wym}. 
Our results, measured on four different lattice spacings and extrapolated to the continuum limit, demonstrate two main effects. First, that QCD matter is described by a negative electric susceptibility at all temperatures.
Second, that the QCD transition, as defined in terms of the Polyakov loop, is shifted to 
higher temperatures as the electric field grows, leading to 
the phase diagram in Fig.~\ref{fig:phasediag}. Furthermore, we showed that lattice simulations employing imaginary electric fields cannot be used to directly assess these aspects due to a singular behavior around $E=0$.

We mention that the susceptibility and the phase diagram are both encoded by the thermal contributions to the real part of the free energy density. These are therefore not impacted by Schwinger pair creation, which is related to the imaginary part of $f$ and is known to be independent of the temperature~\cite{Elmfors:1994fw,Gies:1998vt}.
In other words, the equilibrium charge profile and the polarization of the medium are related to the distribution of thermal charges and not of those created from the vacuum via the Schwinger effect.

Finally we point out that calculations within the PNJL model~\cite{Tavares:2019mvq}, employing the Schwinger propagator, predict the opposite 
picture for the phase diagram as compared to our findings.
Whether the same tendency holds for the Weldon-type regularization within this model, is an open question calling for further study. 
Besides this aspect, the PNJL model is known to miss important gluonic effects in the presence of electromagnetic fields
and fails to correctly describe the phase diagram at $B>0$~\cite{Andersen:2014xxa}.
It would be interesting to see whether improvements that were 
found to correct these shortcomings of the model in the magnetic setting~\cite{Endrodi:2019whh} also work in the $E>0$ case.

\noindent
{\em Acknowledgments}.
This research was funded by the DFG (SFB TRR
211 -- project number 315477589).
The authors are grateful to Andrei Alexandru, Bastian Brandt, David Dudal, Duifje van Egmond and Urko Reinosa for enlightening
discussions.

\appendixsection
\renewcommand{\thesection}{\Alph{section}}
\section{Expansion of the Polyakov loop}
\label{app:ploop}

Here we construct the Taylor expansion of the Polyakov loop expectation value in the background electric field.
We generalize the analogous calculation for the free energy density~\cite{Endrodi:2022wym} to the expectation value $\expv{P_b}$. 

In the presence of the electric field, the equilibrium charge density profile $n(x_1)$ varies in the $x_1$ direction (the coordinate system is chosen so that $-L/2\le x_1\le L/2$). We consider the implications of such an equilibrium
using a homogeneous background field generated by the vector potential $A_0(x_1)=E x_1$, regularized by the finite system size (assuming open boundary conditions). Moreover, the field is assumed to be weak so that the system can be thought of as a collection of subsystems at different $x_1$ with approximately constant density. These are characterized by a canonical free energy density $f$ parameterized by the local density, instead of the usual grand canonical free energy density $\Omega$, parameterized by the chemical potential. The latter is given in terms of the lattice partition function \eqref{eq:pathint} as $\Omega=-T/V\log \Z$. The two free energy definitions are related by a local Legendre transformation~\cite{Endrodi:2022wym},
\be
\label{eq:localLegendre1}
f = \frac{1}{L}\int \dd x_1 \left[ \Omega-\mu \frac{\partial \Omega}{\partial \mu} \right]_{\mu=\bar\mu(x_1)}
\ee
The local chemical potential is fixed by the requirement that diffusion and electric forces cancel, i.e.\ $\bar\mu(x_1)=-eE x_1$. This choice corresponds to a globally neutral system, where the volume average of the chemical potential vanishes.

Including the bare Polyakov loop in the action with a coefficient $\alpha$ and taking the derivative of~\eqref{eq:localLegendre1} with respect to $\alpha$ at $\alpha=0$ results in
\be
\label{eq:localLegendrePloop}
\expv{P_b}^{n} = \frac{1}{L}\int \dd x_1 \left[\expv{P_b}-\mu\, \frac{\partial\! \expv{P_b}}{\partial \mu}\right]_{\mu=\bar\mu(x_1)}\,,
\ee
giving the expectation value of the Polyakov loop in the local canonical ensemble.
Taking the second total derivative of~\eqref{eq:localLegendrePloop} with respect to $eE$, and evaluating it at $E=0$ (implying $\bar\mu=0$), we obtain for~\eqref{eq:Pcalc}
\be
\varphi_E^n = 
\frac{1}{L}\int \dd x_1\left[\varphi_E - \varphi_\mu \cdot x_1^2\right]\,,
\label{eq:varphinE}
\ee
with
\be
\varphi_E =\left.\frac{\partial^2\! \expv{P_b}}{\partial (eE)^2}\right|_{E=0}, \qquad \varphi_\mu=\left.\frac{\partial^2\! \expv{P_b}}{\partial \mu^2}\right|_{\mu=0}\,.
\ee
The Polyakov loop operator $P_b$ does not depend explicitly on the electric field nor on the chemical potential.
The derivatives of $\expv{P_b}$ therefore merely involve the derivative of the 
weight in the path integral~(\ref{eq:pathint}). 

Let us first discuss $\varphi_\mu$.
The chemical potential multiplies the volume integral of $j_4$ in the Euclidean action (before integrating out fermions), therefore 
\be
\varphi_\mu = \int \dd^4 y\, \dd^4 z \left[  
\expv{P_bj_4(y) j_4(z)}-\expv{P_b}\expv{j_4(y) j_4(z)} \right]\,,
\ee
where we used that $\expv{j_4(y)}=0$ due to 
parity symmetry. 
Substituting the integration variable $z$ by $u=z-y$, exploiting the translational invariance of the correlators and using the definition~\eqref{eq:Gdef} of the projected correlator, we arrive at
\be
\varphi_\mu = \frac{V}{T}\int \dd u_1 \left[ \expv{P_b G_{44}(u_1)} - \expv{P_b}\expv{G_{44}(u_1)} \right] \,.
\ee

Next, we turn to $\varphi_E$.
This time, the Euclidean action contains the four-volume integral of $ieA_4(y_1)\cdot j_4(y_1)$ with $A_4(y_1)=-iE\,y_1$.
The second derivative therefore becomes
\be
\varphi_E =\int \dd^4 y\, \dd^4 z \,y_1z_1 \left[  
\expv{P_bj_4(y) j_4(z)}-\expv{P_b}\expv{j_4(y) j_4(z)} \right]\,,
\ee
We proceed by rewriting $y_1z_1 = -(z_1-y_1)^2/2+(y_1^2+z_1^2)/2$ and use that the second term can be replaced by $y_1^2$ as it multiplies a factor that is symmetric under the exchange of $y_1$ and $z_1$ under the integrals.
With the same variable substitution as above, the use of translational invariance of the correlators this time gives
\begin{align}
\varphi_E =  &-
\frac{V}{T} \int \dd u_1 \frac{u_1^2}{2} \left[ \expv{P_b G_{44}(u_1)}-\expv{P_b}\expv{G_{44}(u_1)}\right] \nonumber \\
&+\frac{1}{L}\int \dd y_1 \,y_1^2 \cdot \varphi_\mu\,.
\label{eq:infsingterm}
\end{align}

The second term in~\eqref{eq:infsingterm} is clearly divergent in the thermodynamic limit.
Coming back to~\eqref{eq:varphinE}, we see that this infrared singular term exactly cancels in $\varphi_E^n$, rendering the curvature of the Polyakov loop expectation value finite when evaluated along the equilibrium condition involving the inhomogeneous charge profile.
Finally, employing the $u_1\leftrightarrow -u_1$ symmetry of the $E=0$ system, we end up with Eq.~\eqref{eq:Pcalc} of the main text, involving the second moment $G_{44}^{(2)}$ defined in Eq.~\eqref{eq:chicalc}.

There is one more aspect regarding the dependence of the Polyakov loop on $E$ that deserves mentionting. 
In lattice simulations with constant imaginary electric fields $iE$ at nonzero temperature, the Polyakov loop was observed to develop a local phase proportional to the local vector potential $\arg P_b(x_1)\propto ie E x_1/T$~\cite{Yang:2022zob} (see also the analogous study~\cite{DElia:2016kpz}). This results from the preference of local Polyakov loops towards different center sectors for different $x_1$. Together with the quantization condition for the imaginary electric flux, this corresponds to a topological behavior of the Polyakov loop angle winding around the lattice. Thus, the volume-averaged $P_b$ vanishes in these $i E\neq0$ simulations, as opposed to its nonzero value at $E=0$, showing the singular change of relevant ensembles as the electric field is switched on. Again, we conclude that simulations with homogeneous imaginary electric fields cannot be used for a direct comparison to the $E=0$ system.

\section{Correlators}
\label{app:corr}
Here we discuss the determination of the correlator $G_{44}(x_1)$ and the bare electric susceptibility in more detail. 
The density-density correlator is calculated using $\mathcal{O}(1000)$ 
random sources located on three-dimensional $x_1$-slices of our lattices. 
We take into account both connected and disconnected contributions in the 
two-point function. More details regarding the implementation can be found in~\cite{Bali:2015msa}.
We note that the same two-point function is required, at zero temperature, for the calculation of the hadronic contribution to the muon anomalous magnetic moment, see e.g.\ Ref.~\cite{Meyer:2018til}.

\begin{figure}[b]
 \centering
 \includegraphics[width=8cm]{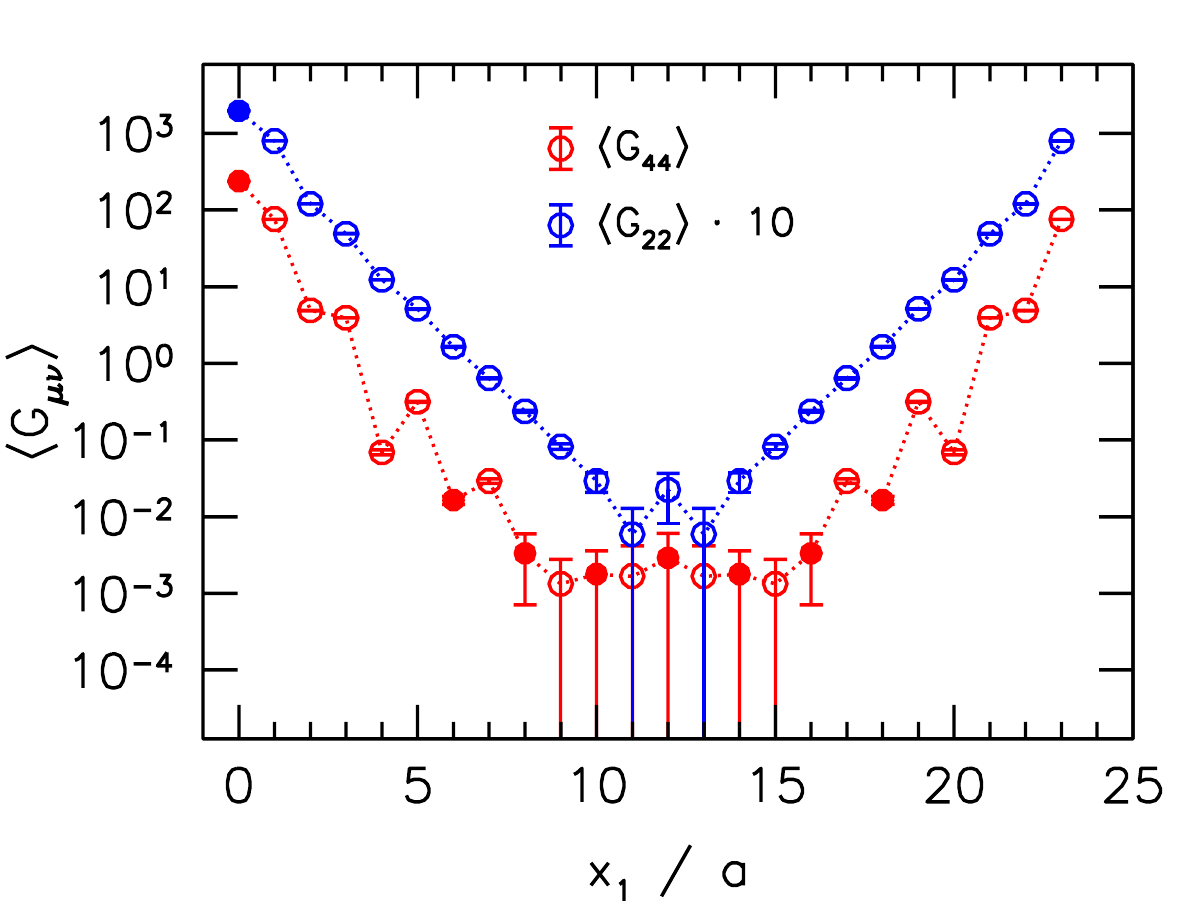} 
 \caption{\label{fig:corr} Current-current $\expv{G_{22}}$ (blue) and density-density $\expv{G_{44}}$
 (red) correlators at $T\approx 176 \textmd{ MeV}$ on our $24^3\times6$ lattices. The former has been multiplied by 10 for better visibility. 
 Filled (open) points indicate positive (negative) values.}
\end{figure}

In Fig.~\ref{fig:corr} we show the zero-momentum projected density-density correlator $\expv{G_{44}}$ as a function of the coordinate
at $T\approx 176 \textmd{ MeV}$. For comparison, the current-current correlator 
$\expv{G_{22}}$, relevant for the magnetic response, is also included. A substantial difference is visible, 
reflecting the absence of Lorentz symmetry at this high temperature. It is interesting to note the systematic oscillation of $\expv{G_{44}(x_1)}$ between even and odd distances -- related to the use of staggered fermions -- which is however absent for $\expv{G_{22}(x_1)}$.

\begin{figure}[t]
 \centering
 \includegraphics[width=8cm]{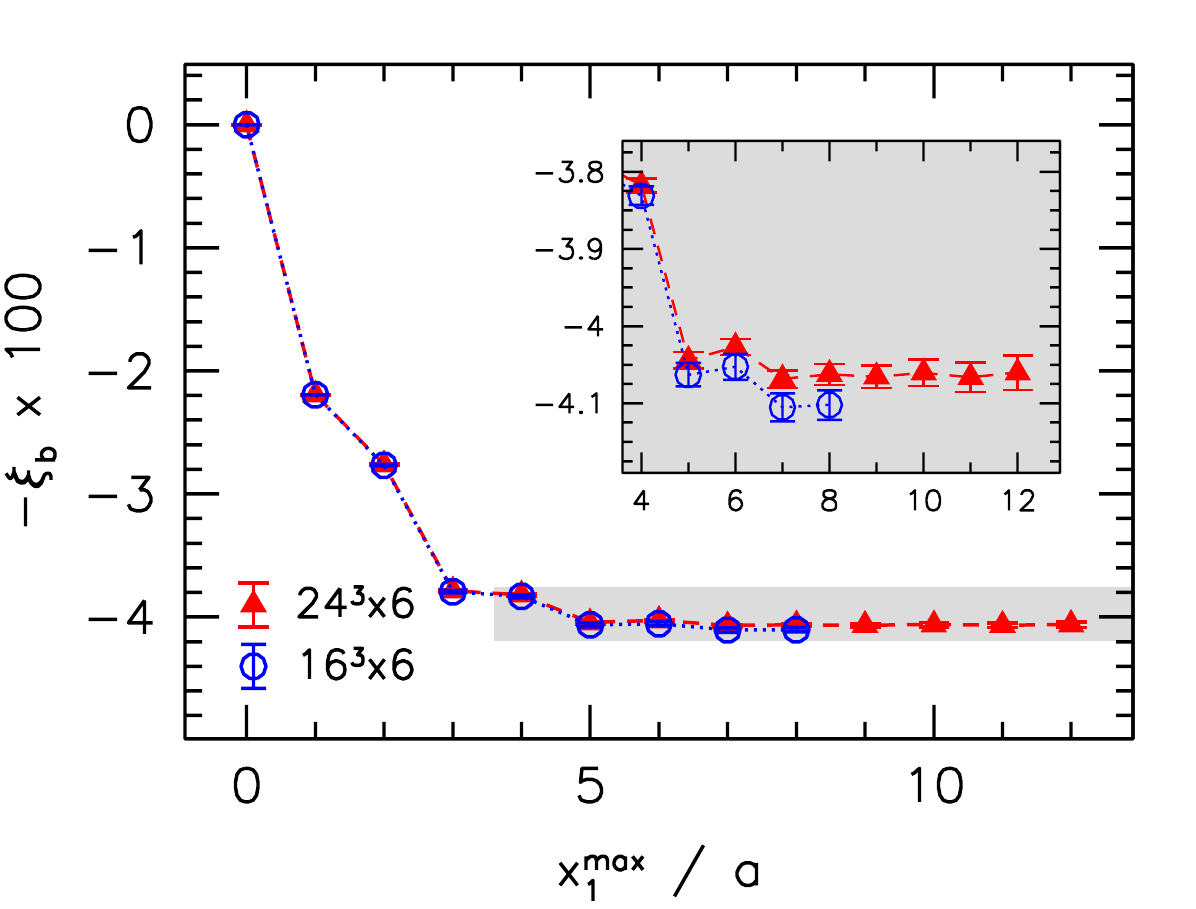} 
 \caption{\label{fig:voldep} Bare electric susceptibility obtained via a truncation of Eq.~\protect\eqref{eq:chicalc} for two different volumes, $24^3\times6$ (red) and $16^3\times6$ (blue). The inset zooms into the region near $x_1^{\rm max}= N_s a/2$.}
\end{figure}

To assess finite volume effects, we consider the convolution~\eqref{eq:chicalc} and truncate it at a distance $x_1^{\rm max}$. The so obtained truncated susceptibility approaches the full susceptibility at $x_1^{\rm max}=L/2$ and is plotted in Fig.~\ref{fig:voldep} for two different volumes.
On the $24^3\times6$ lattice, the plot shows that contributions coming from the middle of the lattice volume are exponentially small, as expected. Moreover, at $x_1^{\rm max}=L/2$, the results obtained on the two volumes agree with each other within errors. The dominant systematic error for the determination of our final result is found to come from the continuum extrapolation, which is discussed in the main text.

\bibliographystyle{utphys}
\bibliography{electric}

\end{document}